\documentclass[a4paper,12pt]{article}

\usepackage{amsmath,amssymb,amscd}
\usepackage{graphicx}

\ifcsname shortitle\endcsname
  \else%
  \fi%

\newcommand{\figref}[1]{\ref{#1}}

\providecommand{\keywords}[1]{\\{} Keywords: #1} 
\providecommand{\shorttitle}[1]{} 
\providecommand{\shortauthorlist}[1]{} 
\providecommand{\name}[1]{#1} 
\providecommand{\address}[1]{\\{}#1} 
\providecommand{\and}{\\{} and } 

\providecommand{\url}[1]{\texttt{#1}} 
\providecommand{\href}[2]{#2} 
\newcommand{\tsedoi}[1]{\url{http://dx.doi.org/#1}}
\newcommand{\tsedoiextra}[1]{} 
\newcommand{\tsearxiv}[1]{\href{http://arXiv.org/abs/#1}{\texttt{arXiv:#1}}}

\begin{document}

\title{Transitive reduction of citation networks}
\shorttitle{Transitive reduction of citation networks} 
\shortauthorlist{J.R.~Clough, J.~Gollings, T.V.~Loach, T.S.~Evans} 

\author{
\name{James~R. Clough, Jamie Gollings, Tamar~V. Loach, \and Tim~S. Evans}
\address{Complexity and Networks group, Imperial College London,\\ South Kensington campus, London, SW7 2AZ, United Kingdom}
}

\maketitle


\begin{abstract}
{In many complex networks the vertices are ordered in time, and edges represent causal connections. We propose methods of analysing such directed acyclic graphs taking into account the constraints of causality and highlighting the causal structure. We illustrate our approach using citation networks formed from academic papers, patents, and US Supreme Court verdicts. We show how transitive reduction reveals fundamental differences in the citation practices of different areas, how it highlights particularly interesting work, and how it can correct for the effect that the age of a document has on its citation count.
Finally, we transitively reduce null models of citation networks with similar degree distributions and show the difference in degree distributions after transitive reduction to illustrate the lack of causal structure in such models.}
\end{abstract}

\section{Introduction}

Citation networks are complex networks that possess a causal structure.  The vertices are documents ordered in time by their date of publication, and the citations from one document to another are represented by directed edges.  However unlike other directed networks, the edges of a citation network are also constrained by causality, the edges must always point backwards in time.  Academic papers, patent documents, and court judgements all form natural citation networks. Citation networks are examples of directed acyclic graphs, which appear in many other contexts: from scheduling problems \cite{BMR88} to theories of the structure of space-time \cite{D06a}. Since causality is such an important feature of any DAG, analysis of such networks should take account of the constraint imposed by time, something standard network measures do not do. Only then can we understand how information flows between agents. We note that the temporal networks we discuss, where time is associated with the vertices, differ from the temporal edge networks discussed elsewhere \cite{HS12}.

\begin{figure}
\centering
\includegraphics[width=0.4\textwidth]{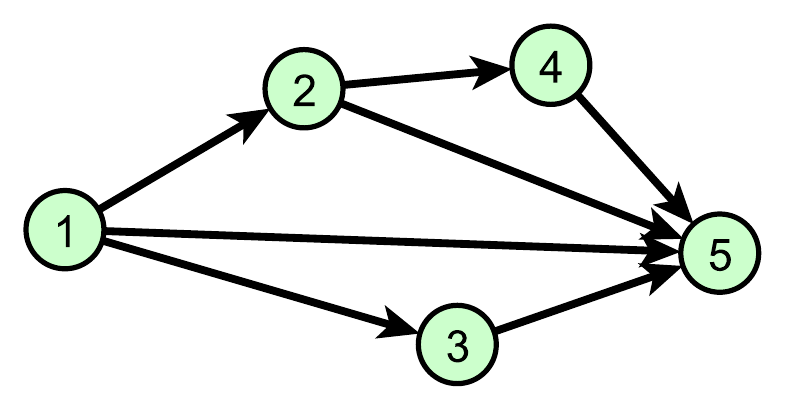}
\hspace*{0.05\textwidth}
\includegraphics[width=0.4\textwidth]{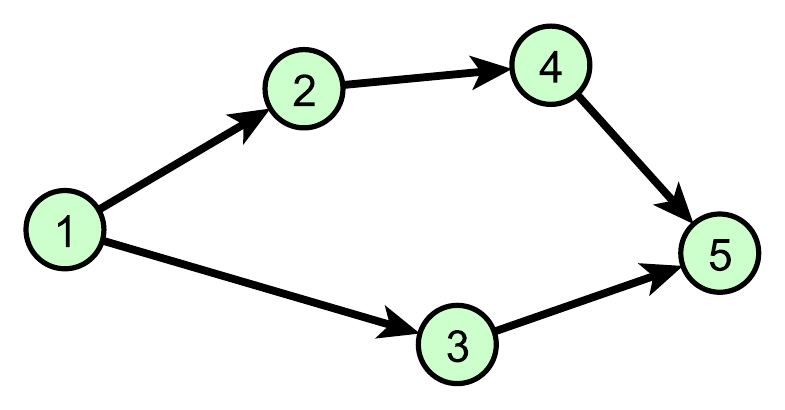}
\caption{Example of a directed acyclic graph (left) and the unique transitive reduction of this network (right).}
\label{ftr}
\end{figure}

Our proposal is to exploit transitive reduction (TR), a well known operation on a directed graph, but only uniquely defined on a directed acyclic graph, because of its causal structure. The transitive reduction of a directed acyclic graph removes all the edges which are unnecessary for the flow of information to be maintained. That is all nodes which were (were not) connected by a path in the original network remain connected (disconnected) in the transitively reduced network, see figure \figref{ftr}. For instance if node A links to B, and B links to C, then transitive reduction removes link A-C, if it exists. This operation may be implemented using standard algorithms \cite{AGU72}.
Once transitive reduction has revealed the fundamental causal skeleton of a network, we may then analyse it using standard network analysis tools.

We will illustrate our approach with just the simplest of network measures, degree. In the context of citation networks, the in-degree is the citation count of a document, the most commonly quoted measure of the importance of a document.  As the number of documents published grows ever faster in all fields (at an exponential rate for academic papers \cite{P61,LI10}) it is increasingly difficult for researchers to find the most important sources. Citation analysis has the potential to alleviate this problem by highlighting documents of high impact and thus of particular interest.  

However, a citation does not always indicate that one piece of work takes useful information from another.  For instance in academia there are many possible motivations for including a paper in a bibliography \cite{B86,MM96}. It has even been argued that most academic citations occur without the author actually having read the paper they are citing \cite{SR03,SR05b}. Likewise court judgements may include references to earlier judgements cited by lawyers but deemed by judges to be of no relevance.  On the other hand, for patents there is a legal requirement to cite `prior art' \cite{JT02} and we will show that citation patterns reflect this different motivation.

Our approach will highlight those academic papers, patents or court decisions which play an essential role in the causal structure of the network.  We suggest that these documents are likely to be particularly important. More generally our methods highlight that the most highly cited papers are not always the most important in terms of the causal structure.
We will also show that our combination of transitive reduction and citation count reveals new overall network features. We believe that this shows that by applying causally aware measures to networks whose vertices are constrained by time, important differences in the network structure and citation behaviour will be revealed which are not seen in traditional network measures.

\section{Data Sets}

Our citation networks are built using documents (individual papers, patents or court judgements) as nodes, with an edge from node A to node B if the document represented by node A has cited B in its text. We also require that each node is associated with a unique time and all edges are directed backwards in time.  In practice, there may be several dates associated with each document: different versions of an academic paper, patents will be filed and granted on different dates. This, along with actual errors in the data, means that it is possible for edges to point in the `wrong' direction, destroying the strict causal structure we require for our work.  In our data sets we use the same characteristic date (specified below for each case) for all the documents.  This ensures almost no edges are acausal but we chose to drop the rare edges which violate the sense of causality defined by our choice of time for the nodes\footnote{The number of unused citations were: arXiv hep-th 1307 (0.4\%), patents 738 (0.005\%), supreme court 540 (0.2\%).}.

Our first example comes from the citation network formed from the bibliographies of academic papers. We use two data sets, each formed from one subsections of the online research paper repository arXiv: hep-ph and hep-th, high energy phenomenology and theory respectively. The citation data we use \cite{KDDarXivData03} is derived only from links between papers within each section, covering papers dated from 1992 (almost the start of the repository) until March 2003. As we find little difference in our analysis for the two separate citation networks we show analysis for hep-th only.\footnote{Equivalent analysis for the hep-ph network is available along with all of our data on figshare\cite {figshare}} The date assigned to each paper is derived from its index on arXiv which gives the order in which papers were first submitted to arXiv.

Secondly we apply our ideas to a patent citation network since patents cite earlier patents. We use data derived from patents registered in the USA between 1975 and 1999\footnote{The data consists of all citations made by patents granted between these dates, so some earlier patents are also included in the citation network but only citations made to them are available} \cite{patent} and used the grant date\footnote{The use of a patent's grant date instead of application date makes little difference to our analysis but was chosen since patent numbers are given sequentially by grant date and so it is this date that provides an easily accessible order \cite{JT02}.} for each patent to define the time associated with a node.

Thirdly, we look at the citation network generated from decisions made by the United States Supreme Court, between 1754 and 2002 \cite{USSC}. In this case the date of the node is the date at which the case was decided.

We will focus on the in-degree $k_\mathrm{in}$ as this is the most common importance measure used for citation networks.  The basic characteristics of these networks are shown in table~\figref{tab:netbasics}.  They are all sparse networks with fat-tailed in-degree distributions and high clustering coefficients.  The largest (US Patents) has nearly four million nodes and fourteen million edges and since our analysis was performed on current basic desktop computers\footnote{Using a desktop computer calculating the Transitive Reduction of the US Patents citation network took a few hours, and the Transitive Reduction of the other, smaller networks takes a few minutes.} it shows that our approach is very usable.
\begin{table}[!t]
  \label{tab:netbasics}
	\centering
	\begin{tabular}{|r||c|c||c|c||c|c|}	
	\hline
	Network & \multicolumn{2}{|c||}{hep-th} & \multicolumn{2}{|c||}{US patents} & \multicolumn{2}{|c|}{USSC} \\
	before/after TR & before & after  & before & after  & before & after  \\ \hline \hline
	$N$ & 27383 & 27383 & 3764094 & 3764094 & 25376 & 25376\\ \hline
	$E$ & 351237 & 62257 & 16510997 & 13996169 & 216198 & 59032\\ \hline
	$C$ & 0.249 & 0 & 0.0757 & 0 & 0.163 & 0\\ \hline
	Mean $k_\mathrm{in}$ & 12.82 & 2.27 & 4.39 & 3.71 & 8.52 & 2.33\\ \hline
	Median $k_\mathrm{in}$ & 4 & 2 & 2 & 2 & 5 & 2\\ \hline
	1st Quartile $k_\mathrm{in}$ & 1 & 1 & 0 & 0 & 0 & 0\\ \hline
	3rd Quartile $k_\mathrm{in}$ & 12 & 3 & 6 & 6 & 11 & 3\\ \hline
	Gini coefficient & 0.729 & 0.481 & 0.684 & 0.670 & 0.620 & 0.510\\ \hline
	\end{tabular}
	\caption{Table of key values for arXiv hep-th, US patent and the US Supreme Court citation networks, before and after Transitive Reduction. \newline
$N$ - Number of nodes \newline
$E$ - Number of edges \newline
$C$ - Clustering coefficient \newline
$k_\mathrm{in}$ - In degree \newline
Gini coefficient refers to the degree distribution
	}
\end{table}

\section{Effect of Transitive Reduction on Data}

\subsection{Transitive Reduction and Network Degree Distribution}

\begin{figure}[h!]
\centering
\includegraphics[width=0.75\textwidth]{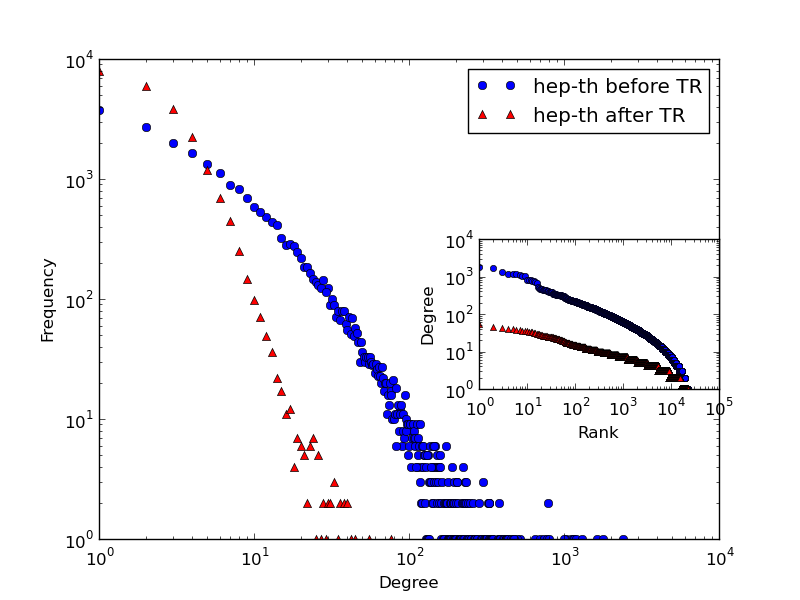}
\caption{TR removes most of the edges from the arXiv network, so the number of low-degree nodes significantly increases and the degree of the high-degree nodes drops. Despite this loss of edges the degree distribution after TR has a similar fat-tailed shape to the original network. The exponent of a power law fit to the tail has increased by ~0.3.}
\label{fig:hep-th_deg_dist}
\end{figure}

In the arXiv citation network around 80\% of the original edges are removed by TR as shown in the table. The clustering coefficient is always zero after TR. Figure \figref{fig:hep-th_deg_dist} shows that TR does not destroy the characteristic fat-tailed degree distribution but the loss under TR is not uniform across all scales: the tail of the degree distribution becomes steeper after TR.  This is clear from figure \figref{fig:hep-th_deg_dist} as well as in the quartile values and Gini coefficients.  In this network papers with more citations lose a larger fraction of their citations under TR.

\begin{figure}[h!]
\centering
\includegraphics[width=0.75\textwidth]{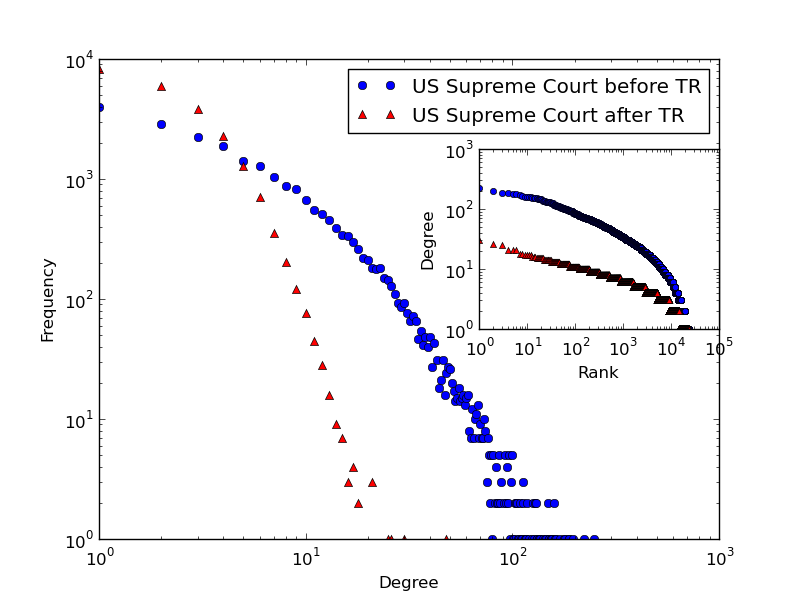}
\caption{The degree distribution for the US Supreme Court citation network has a similar shape before and after TR to the arXiv citation network.}
\label{fig:ussc_deg_dist}
\end{figure}

Applying TR to the USSC (US Supreme Court) citation network gives a similarly sized effect as seen for academic papers, see figure \figref{fig:ussc_deg_dist}, and here around 73\% of edges are found to be unnecessary for the causal relationships.

\begin{figure}[h!]
\centering
\includegraphics[width=0.75\textwidth]{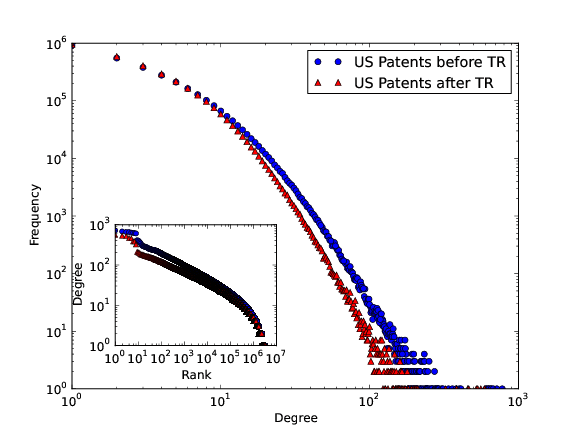}
\caption{TR only removes a small fraction of edges in the Patent citation network illustrating that its causal structure differs significantly from the arXiv, or USSC citation networks.}
\label{fig:patent_deg_dist}
\end{figure}

On the other hand, the US patent network TR has a much smaller effect, removing only around 15\% of the edges in the network and figure \figref{fig:patent_deg_dist} shows the change in degree distribution. Despite the degree distributions of the arXiv citation network, the USSC network and the patent network having a similar shape, their different causal structures are revealed by the way that TR affects them.
The post-TR degree distribution reveals that the citation behaviour of academics and judges is very different from that of inventors, patent lawyers and patent examiners.  This reflects what we understand about the citations processes in each area.

Academics cite papers for a variety of reasons \cite{B86,MM96}.  Old and well known works are often cited in a paper because they contain essential ideas, but could also just be cited as standard for that field despite having no direct influence on the current work. Our results suggest that referees and editors of academic journals are not providing an efficient independent check that authors' citations are necessary.  Indeed there are clear incentives in the system for authors, referees and editors to add citations to their own publications regardless of the need to do so, for instance \cite{EHK12}.

Likewise a highly cited court judgement may be one which sets out new rules or principles.  However it has also been suggested that citations in judgements are influenced by other factors: the reputation of a judge writing the earlier document, the clarity rather than originality of an previous judgement.  Sometimes several judgments making the same point are quoted to indicate a weight of evidence behind a conclusion. It may even happen that an earlier judgement is mentioned by lawyers in a case yet its relevance is dismissed by the judges; such a negative assessment will still produce a citation.

On the other hand, the patent citation process is quite different since patent citations serve a particular legal function \cite{JT02}. When patent A cites patent B it implies that patent B contains pre-existing `prior art' over which A has no claim. There is no need, or incentive for an inventor or patent lawyer to unnecessarily cite the same information many times, as the existence of information in just one patent is enough designate it as prior art. Independent patent examiners enforce this legal obligation to cite pre-existing work and have no need to cite the same information more than once.  
We should therefore expect to see many citations in the arXiv and USSC citation networks are not required for the flow of information to exist, while the patent citation network will have few such unnecessary citations.  This is something that the change in degree distribution after TR illustrates.

\subsection{Effect of Transitive Reduction on Individual Nodes}

Interesting behaviour can also be seen by looking at how the in-degree (citation count) of individual nodes changes under TR.

\begin{figure}[h!]
\centering
\includegraphics[width=0.6\textwidth]{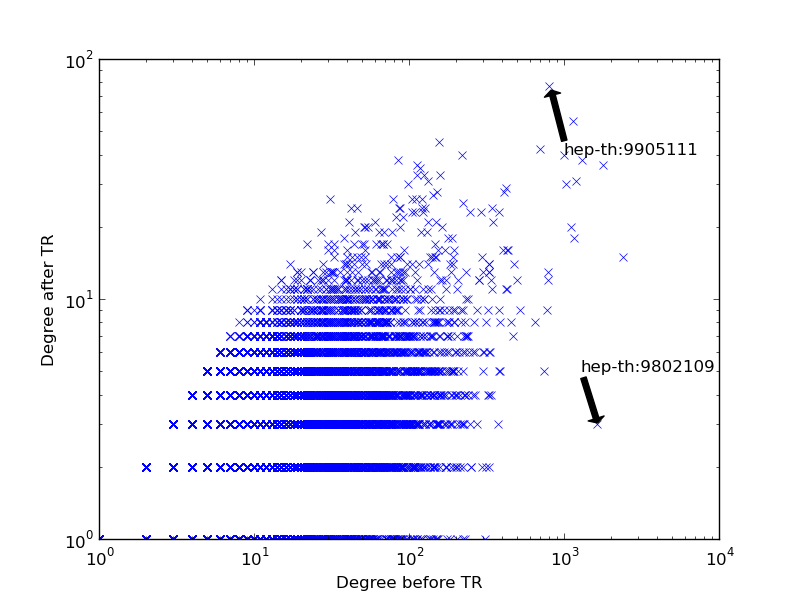}
\caption{Paper 9802109 originally has 1641 citations, making it the 3rd most popular paper in the network, but after transitive reduction it loses 99.8\% of its citations, and has only 3 left. Paper 9905111 begins with 806 citations, and is around the 10th most popular paper in the network but after transitive reduction it retains 77, a loss of  90.5\%.}
\label{fig:hep-th_deg_change}
\end{figure}

\begin{figure}[h!]
\centering
\includegraphics[width=0.6\textwidth]{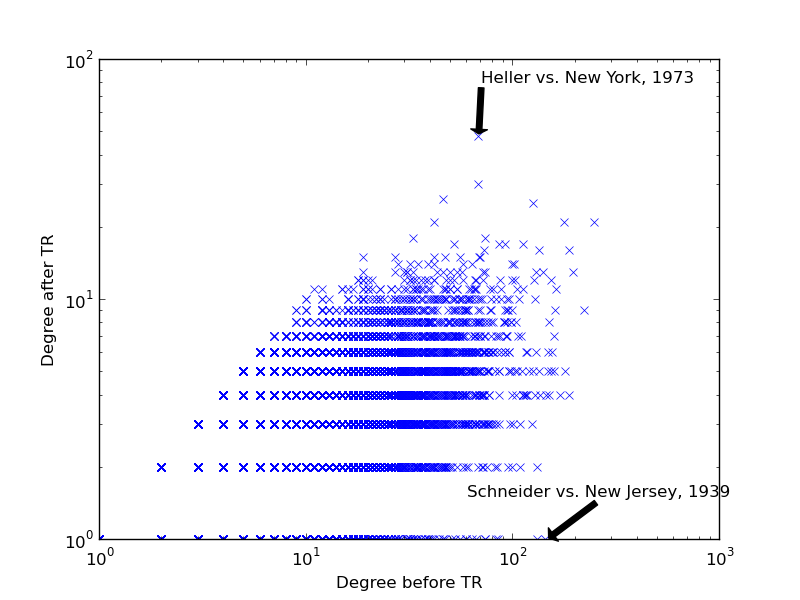}
\caption{As with the degree distributions, the range of degree change after TR is large in the USSC network. Many highly cited cases become very poorly cited after TR.}
\label{fig:ussc_deg_change}
\end{figure}

In the case of the arXiv citation network and the US Supreme Court citation network, there are large variations in behaviour for individual papers as figures \figref{fig:hep-th_deg_change} and \figref{fig:ussc_deg_change} show.

For instance, in the arXiv network, paper hep-th/9802109 (`Gauge Theory Correlators from Non-Critical String Theory' by Gubsner et al.) was cited by 1641 papers in the network, but only three citations remained after TR. This means that of those 1641 papers, three were special in that the other 1638 all also connected to this one via one of those three papers. Anyone who took information from this hep-th/9802109 also took information from one of hep-th/9802150, hep-th/9906004, or hep-th/9902130. This fact is made explicit by transitive reduction.

On the other hand, paper hep-th/9905111 (`Large N Field Theories, String Theory and Gravity' by Aharony et al.) begins with a similar number of citations, 806, yet after TR it retains 77 of these.  This means that there were 77 papers in this network which cited hep-th/9905111 but who did not cite each other. It seems information taken from this paper was utilised more diversely.

A similarly wide range of citation count changes occurs in the Supreme Court citation network. The case `Schneider vs. New Jersey (1939)' has 144 citations but after TR falls to just one, and `Stromberg vs. California (1931)' also falls from 132 citations to just one. Conversely, the case `Heller vs. New York (1973)' goes from having 68 citations, to 48 after TR becoming the most cited case in the whole network. The second most cited case, after TR, is `Hamling vs. United States', going from 68 to 38 citations. Wikipedia lists hundreds of Supreme Court cases but these two do not appear famous enough to make that list, even though our analysis shows that they have a very significant place in the causal structure of that citation network.

In the US patent citation network there is far less variation between individual nodes of the effect of TR. Most patents lose just a small fraction of their citations. Again, this suggests a different kind of citation behaviour in general, where highly cited patents are mostly cited by a diverse group who do not cite each other.
Despite that general trend, there are still outliers which lose a significant fraction of citations after TR, illustrating the use of TR to highlight nodes whose citations form a more narrow or broad group.

\begin{figure}[h!]
\centering
\includegraphics[width=0.6\textwidth]{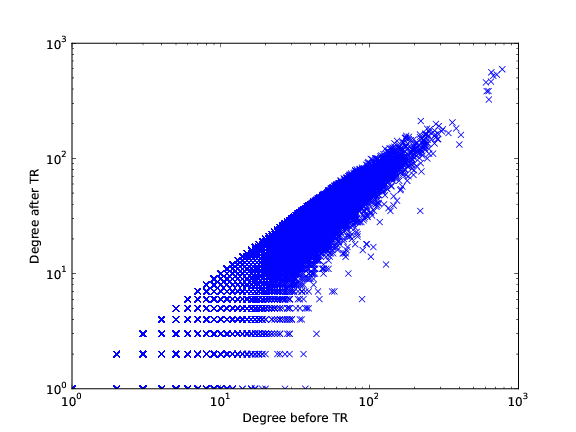}
\caption{Unlike the arXiv citation network, in the patent citation network the fraction of edges lost after TR is similar for high degree and low degree nodes and so degree after TR is highly correlated with original degree. Individual nodes of interest can still be highlighted. A patent on `Arylpiperazinylalkoxy derivatives of cyclic imides (1988)' with a degree of 67 lost no edges after TR, showing diverse citation behaviour, but a patent on `Improvement in the manufacture of artificial stone' dropped from degree 37 to degree 1 showing narrow behaviour.}
\label{fig:patent_deg_change}
\end{figure}

\subsection{Using Transitive Reduction to correct citation count for paper age}

\begin{figure}[h!]
\centering
\includegraphics[width=0.6\textwidth]{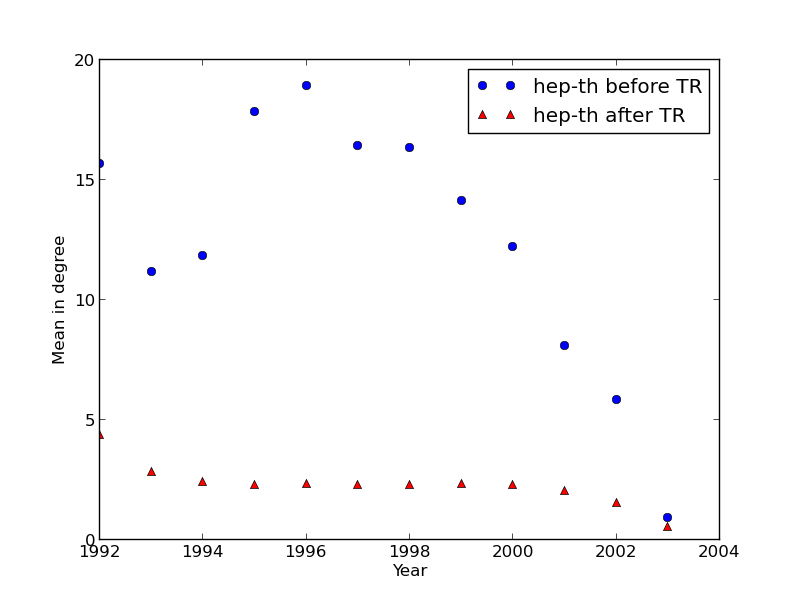}
\caption{In the original citation network older papers have many more citations than newer papers. The citation count is not based solely on scientific worth but on the age of the paper since citations accrue over time as more scientists read the paper. After transitive reduction this effect is greatly reduced. We discount the results for the first three years 1992 - 1994 of our data as these were the first years of  arXiv grew rapidly.}
\label{fig:hep-th_years}
\end{figure}

Looking at figure \figref{fig:hep-th_years}, we see that the mean citation counts of academic papers increases as they get older, a well known pattern.\footnote{Although this is not true for the first three years of the citation network, probably because papers most often cite recently published work and in the early years of the arXiv there were many fewer papers uploaded that were available to cite, and our data only captures citations between papers which are both on the arXiv.}

After TR the average number of citations rises slowly, but reaches a plateau after only three or so years.  One immediate implication comes from this stability in the average TR citation count over time: it makes sense to use the \emph{post}-TR in-degree as the measure of a paper's impact when comparing papers with different publication dates.

It is not surprising that older papers lose a larger fraction of their citations during TR. They have more citations with a large age difference and these links are more likely to be cut by TR as there are many possible causal paths between the two nodes besides their direct link.

The difference before and after TR in the temporal behaviour of average citation counts, figure \figref{fig:hep-th_years}, and in the large variation of the citation counts of individual papers, figure \figref{fig:hep-th_deg_change}, suggests the post-TR count is revealing a different type of citation behaviour in academia. Typically someone who cites an academic paper shortly after the older paper was published is building directly on fresh research and TR retains such links.  However if a citation refers to an old standard paper many years after its publication it is likely that the original ideas of the older paper were developed by others, who themselves cited this standard paper.  Later researchers may well include both in their bibliography, even though the older publication probably had no direct influence on the recent publication. TR discounts such behaviour by removing these links. Thus, by revealing the minimal underlying causal structure of the citation network, the citation count after TR highlights a different type of impact than that shown by traditional citation counts.

There is support for our idea that TR may highlight citations which show genuine transfer of inspiration and innovative ideas. The statistical analysis of typographical errors conducted by Simkin and Roychowdhury \cite{SR03,SR05b} suggests that around 78\% of academic citations occur without the author having even read the paper they cite. The model used to arrive at this figure assumes that an author reads one paper, cites it, then goes on to cite older papers (without reading them), simply copying the citation (along with any errors) from the first paper's bibliography. Such copied links are precisely the ones TR removes.  Conversely, the edges that are not removed during TR are ones where there is no intermediary paper with a bibliography to copy, so they are far more likely to correspond to a citation that was actually read. Interestingly TR on the two arXiv citation networks leaves roughly 20\% of the original citations, matching the claim of Simkin and Roychowdhury that only 22\% of papers have been read at all \cite{SR03,SR05b}. The similarity between these two figures is strong evidence that TR has vastly increased the probability that a citation in the TR network corresponds to a paper that has been read. It would be extreme to claim that \emph{all} of the deleted citations were bad citations copied from bibliographies, but we have deleted almost all of these bad citations, plus some unknown number of good, paper-reading citations.

Another aspect is that for an academic paper a high citation count after TR indicates that the influence of a paper has rippled widely through many fields, as papers within different fields are unlikely to be closely linked in the network.  A low citation count after TR means the influence of that paper is confined to a narrow set of papers that largely refer to each other.  This immediately suggests that one use of a TR citation count is to highlight interdisciplinary papers and so it should be used to recommend papers to researchers looking outside their usual area of expertise.

\begin{figure}[h!]
\centering
\includegraphics[width=0.6\textwidth]{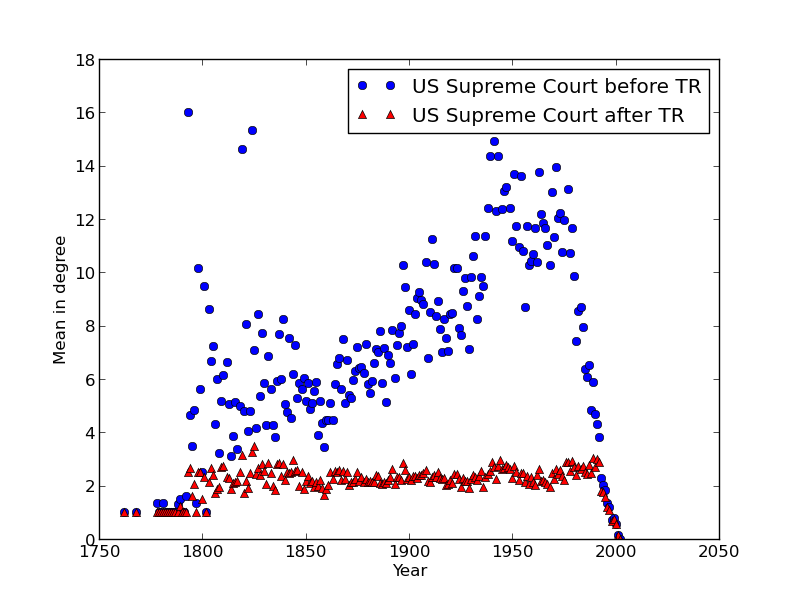}
\caption{In the original US Supreme Court citation network, there are few judgements in the early years of the court, but some are highly cited as they set important precedents early in the court's history. After this initial period the average citation count per year is lower but slowly grows, peaking in recent years, before rapidly dropping off as new judgements are yet to be cited at all. After TR the citation count per year is almost constant until the last few years of the data.}
\label{fig:ussc_years}
\end{figure}

The data on citation of previous judgements in the US Supreme Court shows a dependence on the age of the judgement as shown in figure~\ref{fig:ussc_years}.  For recent judgements over the last fifty years or so, the older the judgement is, the more citation it gathers, just as is seen with academic papers.  However after that the average number of citations falls away steadily away though with a fair amount of scatter. This is to be contrasted with the plateau in citation number on long time scales often seen in academic citations. However it is the effect after TR that is of interest here and what we see is that the essential causal structure revealed by TR shows a similar pattern for both.  For supreme court judgements, only judgements from the last ten years are essential as all older judgements which are mentioned are referred to from those given in the last ten years.

\begin{figure}[h!]
\centering
\includegraphics[width=0.6\textwidth]{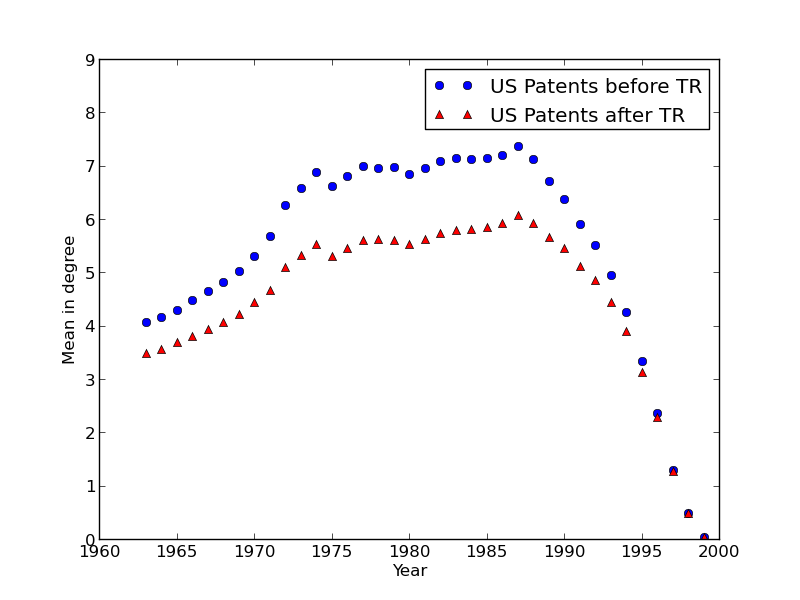}
\caption{In the original US patent citation network citation count shows a period of initial growth from the 1960s through to ~1975, and then plateaus until ~1990 where it rapidly drops off, as very new patents are not old enough to be cited. After Transitive Reduction, this pattern is unchanged suggesting a different kind of citation behaviour to the arXiv citation network.}
\label{fig:patent_years}
\end{figure}

In the US patent network, many fewer edges are lost after TR and we do not see a large difference in behaviour for older and newer patents. Once they are old enough to start being cited (figure \figref{fig:patent_years} suggests this period is around 8 years) we do not observe older patents losing a larger fraction of their citations under TR than younger ones.

Our TR process is again showing fundamental differences in the in citation practice of academics (who generate the arXiv citation network), inventors and  patent lawyers (who generate the US patent citation network) and the judges writing US Supreme Court judgement. Academics cite for a variety of reasons, some of which encourage them to cite old, influential papers in their field even if their current work has not been directly influenced by that paper. Judges will try to back up their conclusions by referring to as many older judgements as they can, even if each previous judgement reiterates the same point. Thus both academics and judges cite earlier documents which were not essential to explain their current work, and they could have confined themselves to documents produced over the last three or ten years respectively. The post-TR distribution of citation count against document age clearly reveals the similarity in the practices of these two distinct fields.

On the other hand our analysis shows clearly that inventors and patent lawyers use citations in a completely different way. They do not write a patent because they want other inventors to read and cite it, nor do they want to highlight where the same idea appeared in earlier patents. Rather they are only interesting in citing others because they are legally obliged to cite prior art. It is a necessary part of the process of getting their own patent. So inventors have no reason to cite anything which is not a direct influence on their own invention.  TR reveals this in the way that far few patents are removed under TR and the way that the post-TR citation count tracks the full data whatever the age of the document.

\section{Comparison with Null Models}

It is often difficult to assess whether a model builds a complex network that is similar to one found in the real world. Since these networks are large and not regular in structure, they contain masses of data which we need to prioritise and condense into useful information. Analysing the degree distribution is one way of testing a model, so a model which purports to create networks similar to actual citation networks should have a similar degree distribution. We argue that the degree distribution after TR is also an important test of similarity in networks where the nodes are ordered in time, and the causal structure is important. This is because TR is a particular and unique operation which highlights certain, particularly important edges in the network: those which are necessary for its causal structure.

\begin{figure}[h!]
\centering
\includegraphics[width=0.75\textwidth]{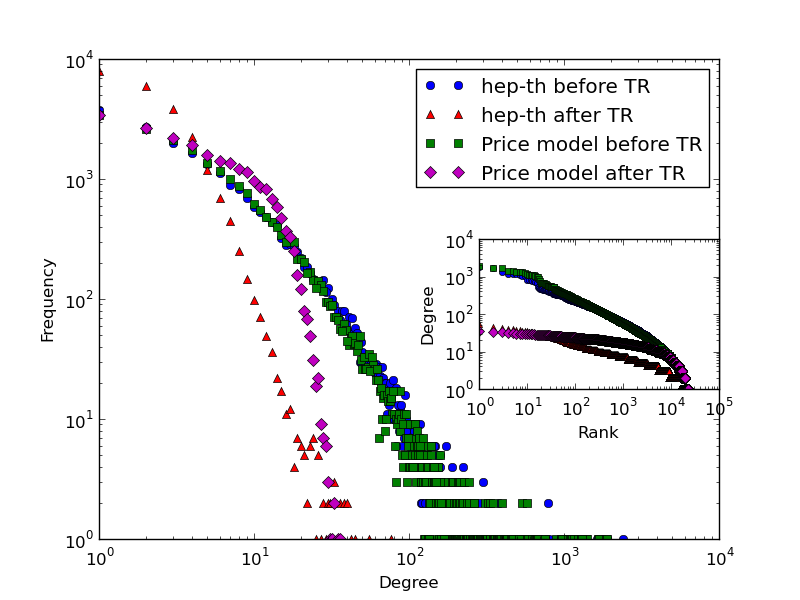}
\caption{A directed acyclic graph made using a cumulative advantage model has a similar degree distribution to the arXiv network before TR. Before TR the degree distributions look similar (see the circles and squares). After TR (triangles and diamonds) the arXiv network is clearly distinguishable from the model illustrating that they have different causal structure.}
\label{fig:BA_TR}
\end{figure}

To illustrate this idea we use a mixture of cumulative advantage of the original Price model of citations \cite{P61} (preferential attachment in \cite{BA99}) along with random vertex attachment, to create a citation network of the same size and similar degree distribution as the hep-th data, but without any of the other structure present in a citation network.  However, figure \figref{fig:BA_TR} illustrates that after TR, the cumulative advantage model no longer has a degree distribution similar to that of the citation network. Clearly the Price model has a different causal structure.

So we have shown that comparing a null model with a real network after transitive reduction reveals discrepancies (even in simple measures such as degree distribution) which are not seen when comparing the pre-TR networks. We suggest that applying this process to other models is a good test to see if the model captures the essential causal structure in the data.

\begin{figure}
\centering
\includegraphics[width=0.4\textwidth]{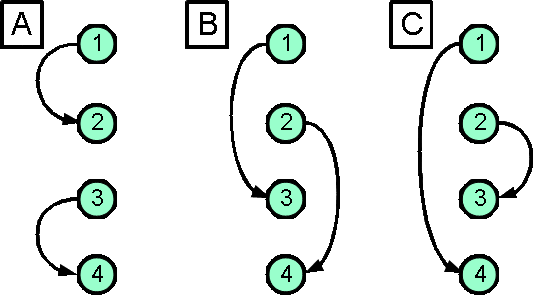}
\caption{The only three arrangements of two directed edges between four distinct vertices in a directed acyclic graph where vertex index specifies their order. Under a rewiring where both the in- and out-degree are maintained, arrangement A cannot be changed, whilst B and C can only be interchanged.  Other rewirings that maintain degree distribution are possible, but the edge directions would not respect the order of the vertices.}
\label{fig:rewiring}
\end{figure}

Another standard way to obtain a null model is to randomise the edges in the data set of interest.
A common approach is to randomly rewire the edges of a network to remove structure while maintaining the degree distribution. This can be done by taking a pair of edges [A, X] and [B, Y] and swapping them to [A, Y] and [B, X], a well known procedure on networks. However, when vertices are ordered by causality this rewiring is also constrained. Figure \figref{fig:rewiring} illustrates that there are three possible configurations for a pair of edges in an ordered directed acyclic graph. Configuration A cannot be rewired without violating the ordering constraint, and configurations B and C can only be rewired into each other. This operation is equivalent to constructing a random directed acyclic graph with a fixed degree sequence \cite{KN09}.

\begin{figure}[h!]
\centering
\includegraphics[width=0.75\textwidth]{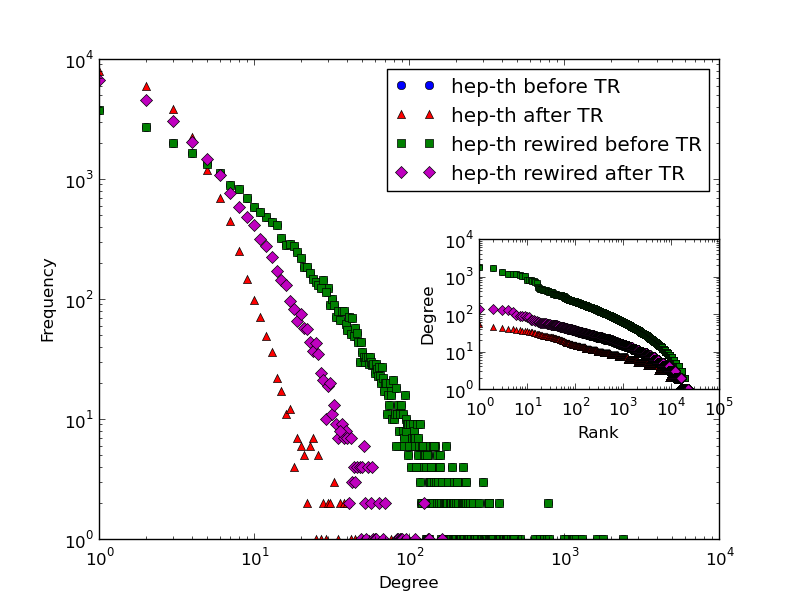}
\caption{Rewiring the edges of the citation network does not change its degree distribution but does change its structure. This change is observable in the degree distributions after TR. The rewired network loses fewer edges by TR as there are fewer highly clustered nodes and edges forming a triangle, which are removed.}
\label{fig:rewiring_deg_dist}
\end{figure}

If a citation network is rewired in such a way as to maintain both the time ordering of the nodes and the original degree distribution, TR again reveals the extent to which structure is maintained as shown in figure \figref{fig:rewiring_deg_dist}. After rewiring\footnote{We randomly rewired edges of the hep-th citation network 10 million times, so each edge has been swapped approximately 30 times. This should completely remove any structure beyond the degree distribution, and the node ordering.}, TR is seen to remove fewer edges in the edge-rewired network, showing that randomly rewired edges are less likely to be implied by transitivity than those edges representing true citation behaviour. However, the similarity between the shapes of the degree distributions of the networks after TR with and without rewiring suggests that much of the structure of the network is retained even after every edge is rewired, because of the constraint of time ordering. This would not be the case had the rewiring process been applied without considering the ordering of vertices. Notably, there is significantly more structure in a completely rewired citation network than a cumulative advantage model, as revealed by TR.

\section{Conclusions}

Our fundamental message is that when analysing a network with constraints, it is essential to use measures which take these limitations into account. This is well known for networks embedded in space \cite{B11a,EEBL11} but when the constraint is a time ordering on the vertices, we need new analysis tools such as TR and edge rewiring that accounts for causality as suggested here. The TR of a causal networks reveals important information on the network structure which is not seen by other characterisations or measurements of networks.

We have illustrated this principle by looking at citation networks, focussing on the simplest measure, the in-degree or citation count of documents. We have shown that the behaviour of the degree before and after TR shows large differences for the academic paper and court judgement networks, and a much smaller change for the patents network. This is also reflected in the way that average citation count after TR varies with document age. With both of these measurements, TR is showing that academics and judges cite a large number of documents whose existence could be deduced from the small number (around 20\%) of recent documents which TR shows is essential for the causal structure.  It is not possible to say \emph{all} the citations removed by TR provided no directed influence, however it does seem that the process of innovation is most strongly stimulated by the few documents TR highlights. For academic papers, this fraction matches  the proportion of citations found to be important for a publication using other methods. TR also highlights the fundamental difference between the citation practices of patents and that of academics and judges.

While the data on average citation counts is illuminating, looking at the citation count of individual documents before and after TR highlights important variations.  For academic papers we have interpreted a higher than expected citation count after TR as an indication that results in a paper were used across a wide number of fields.  As such it could be a useful indicator when searching for `interesting' papers to study, especially when looking for papers outside one's own field that may have cross-disciplinary interest.

Our conclusions are reinforced by the application of our techniques to null models. We again note the need to take time into account, either through the Price model's explicit causal mechanism, or through our modified edge swapping scheme. TR reveals that these models do not capture the fundamental causal structure revealed by TR in the data.

The plots and data used in this paper are publicly available \cite{figshare}.

\section*{Acknowledgements}

We would like to thank K.Christensen, A.Hughes, E.Viegas and R.Wyatt for useful discussions.

\end{document}